\documentclass[12pt,notitlepage]{article}%

\usepackage[comma]{natbib}
\usepackage{url}
\usepackage{geometry}
\usepackage{amsmath}
\usepackage{amsfonts}
\usepackage{amssymb}
\usepackage{graphicx}%
\providecommand{\U}[1]{\protect\rule{.1in}{.1in}}
\begin{document}

\title{The geometry of information coding in correlated neural populations}
\author{Rava Azeredo da Silveira$^{1,2,3,4}$ and Fred Rieke$^5$\\
{\small $^1$ Department of Physics, Ecole Normale Sup\'{e}rieure}\\
{\small $^2$ Laboratoire de Physique de l'ENS, Universit\'e PSL, CNRS, Sorbonne Universit\'e, Universit\'e de Paris}\\
{\small $^3$ Institute of Molecular and Clinical Ophthalmology Basel}\\
{\small $^4$ Faculty of Science, University of Basel}\\
{\small $^5$ Department of Physiology and Biophysics, University of Washington}}


\maketitle

\begin{abstract}
Neurons in the brain represent information in their collective activity. The
fidelity of this neural population code depends on whether and how variability
in the response of one neuron is shared with other neurons. Two decades of
studies have investigated the influence of these noise correlations on the
properties of neural coding. We provide an overview of the theoretical
developments on the topic. Using simple, qualitative and general 
arguments, we discuss, categorize, and relate the various published results.
We emphasize the relevance of the fine structure of noise correlation, and we
present a new approach to the issue. Throughout we emphasize a geometrical
picture of how noise correlations impact the neural code.

\end{abstract}

\noindent
When citing this paper, please use the following:
Azeredo da Silveira R, Rieke F. 2020. The geometry of information coding in correlated neural populations. Annu. Rev. Neurosci. Submitted. DOI: 10.1146/annurev-neuro-120320-082744


\section{Introduction}

The quantitative study of information processing by neurons was born from
investigations that correlated neural responses with parameters that
characterize an `external event' such as a physical stimulus or a motor action
\citep{hubel1995eye}. Responses of single neurons to simple stimuli have
revealed many key properties of coding. These are often summarized in the form
of receptive fields or, equivalently, tuning curves. Except in rare
cases, however, physical stimuli and motor actions are coded, or represented,
in the activity of an entire population of neurons. How, then, are 
rich signals represented in the collective activity of many neurons?

The issue of noise is central. Responses of a single, ideal, noiseless neuron
can encode an infinite amount of information. By contrast, a real, noisy
neuron disposes of a finite bandwidth. In a population of neurons, the noise
in each individual neuron can
be reduced by averaging. This is the simplest view on population coding: the
population enhances the representative signal by averaging out the noise. But
there are two features of population coding that make it a richer problem.
First, physiological properties differ among cells, so that different neurons
represent different aspects of the stimulus. Second, noise in the responses of
individual neurons is correlated and, hence, its impact on information coding
has to be considered collectively, not one cell at the time. These two aspects
of the problem are intimately related: neurons acquire diverse properties
because of the specificity of the connections they make to other neurons, and
this also shapes the correlation in the noise. For example, divergence of
common inputs may permit parallel channels to each encode a different aspect
of the inputs but may also result in strong noise correlations between the
channels. More generally, the architecture of neural circuits shapes the
structures of both signal and noise.

A great deal of research in quantitative neuroscience attempts to relate the
geometry or statistics of neural responses to sensory stimuli or task
parameters. This problem is difficult because it is high-dimensional, as both
signal and noise are specified by a number of possible patterns that grows
exponentially with the number of neurons. Statistical physics exemplifies a
possible way to tame this complexity: phenomena such as phase transitions and
superconductivity were explained by identifying the collective variables most
relevant to the dynamics of measured quantities. Once these relevant
variables---specific combinations of the microscopic variables---were
identified, phenomena of interest could be explained simply in terms of energy
stored in the collective variables or of fluctuations thereof. The
understanding of neural coding would similarly benefit from the identification
of analogous collective variables. Indeed, a great deal of effort is expended
today to develop methods that can extract `low-dimensional' or `latent'
variables from recordings of neural populations. An important consideration,
here, is the need to consider the structure of the average population activity
(e.g., in response to a set of stimuli) as well as the statistics of the
variability about this average, and how the two relate.

In the past two decades, progress on understanding how coding depends on the
geometry of signal and noise has been promoted by a simplifying choice,
namely, a focus on pairwise correlations. These, unlike higher-order
statistical moments, can be measured within the duration of typical neural
recordings. Many neural systems exhibit non-negligible pairwise correlations
\citep{hatsopoulos_donoghue_1998,mastronarde_1989,ozden_wang_2008,perkel_moore_1967,sasaki_llinas_1989,zohary_newsome_1994,shlens_chichilnisky_2008,usrey_reid_1999,vaadia_aertsen_1995,bair_newsome_2001,fiser_weliky_2004,kohn_smith_2005,smith_kohn_2008,lee_georgopoulos_1998,ecker_tolias_2010,graf2011decoding,goris2014partitioning,lin2015nature}.
Early on, also, pairwise noise correlations were hailed as relevant to coding
and behavior:\ the limits they imposed on noise reduction by averaging across
neurons was hypothesized to account for the surprisingly similar detection
thresholds of small populations of neurons and entire animals
\citep{zohary_newsome_1994,bair_newsome_2001}.

These experimental findings and some other early investigations
\citep{johnson_1980,vogels_1990,oram_sengpiel_1998} motivated a series of
studies that set heuristic arguments on firm bases and expanded on them, using
detailed models of population coding
\citep{abbott_dayan_1999,sompolinsky_shamir_2001,wilke_eurich_2002,romo_salinas_2003,golledge_young_2003,averbeck_lee_2003,shamir_sompolinsky_2004,shamir_sompolinsky_2006,averbeck_lee_2006,averbeck_pouget_2006,josic_delarocha_2009}
and general information theoretic arguments
\citep{panzeri_rolls_1999,pola_panzeri_2003}. In addition to elucidating how
noise correlation can limit coding, some of the early work
\citep{abbott_dayan_1999,wilke_eurich_2002} raised the possibility that noise
correlation need not always harm coding. More recent investigations
\citep{ecker2011effect,hu2013sign,silveira2013high,moreno2014information,franke2016structures,zylberberg2016direction}
expanded the panorama of possible scenarios by showing that noise correlation
can be harmless or appreciably beneficial to the neural code. The key here was
the consideration of the fine structure of correlation, beyond its magnitude.
Indeed, analyses of retinal
\citep{franke2016structures,zylberberg2016direction} and cortical recordings
\citep{averbeck_lee_2004,averbeck_lee_2006,montani2007role,graf2011decoding,lin2015nature,montijn2016population}
have illustrated the beneficial impact of specific structures of noise
correlations on coding.

Here, we review theoretical developments on neural population coding in the
presence of correlated noise. We provide an overview of the topic that
combines heuristic arguments, the study of simple models, and general
mathematical statements. To ensure a formal unity, we focus primarily upon the
mutual (Shannon) information as a means to quantify the neural code, and we
comment on its relations with other, related quantities. Section 2 introduces
the problem of neural population coding. Section 3 reviews early, heuristic
arguments that pointed to a potentially detrimental role of noise correlation
in coding. Section 4 presents a general, qualitative argument that encompasses
more recent models, and delineates the conditions under which noise
correlations are detrimental, inconsequential, or beneficial. Section 5
presents a model-independent point of view of the problem by expressing the
mutual information in a form that delineates the role of different types of
correlation. Section 6 examines the coding problem from a geometrical point of
view that complements and further clarifies the results described in earlier
sections and in the recent literature.

\section{The problem of neural population coding}

Sensory stimuli are coded in the activity of populations of neurons. One of
the fundamental problems in neuroscience is that of elucidating the nature of
this code; this problem can be divided into two parts. On the encoding side,
we would like to know what properties of the population activity are relevant
to the representation of information, and how these properties are manipulated
by the brain. On the decoding side, we would like to identify the mathematical
operation that retrieves a physical stimulus (or some feature of it) from the
output of a population of neurons. We can then ask also how such a
mathematical operation is implemented by neurons. Here, we are concerned
exclusively with the encoding side of the problem. Earlier reviews (see, e.g.,
\citep{averbeck_pouget_2006}) discuss the impact of correlations on decoding.

Population coding is a much richer problem than single-cell coding because it
is high dimensional. The number of population states grows exponentially with
the number of neurons, allowing for combinatorial codes. This is true even for
noiseless neurons, as cells come in different functional (and genetic) types
and even cells of a given type present physiological variability. The
situation is further complicated by the fact that neurons are noisy: a given
physical stimulus can elicit one of a number of population activity patterns.
(We are not making any philosophical statement about noise as a sort of
fundamental randomness. Instead, we refer to noise in a procedural way: for
example, we say that the neural response is noisy if it varies from one trial
to the next of an identical stimulus. This variability may result from
biochemical stochasticity, but it may also reflect the purely deterministic
dynamics of a complex system, such as interference between coding of the
visual stimulus with other, unrelated neural activity elicited by 
other stimuli or internal processing.) The mean population response to a
sensory stimulus and its variability are given by the joint statistics of the
firing of neurons. Thus, the fundamental problem of neural population encoding
amounts to asking how information about a physical stimulus is represented by
this complicated mathematical object. We say `information
about the physical stimulus' rather than specifically the stimulus itself
because a neural population may represent properties associated with the
stimulus, such as one of its attributes, a hidden event that may have caused
the stimulus, or even a `meta-property' of the stimulus such as the
probability with which it occurs in a specific environment.

Following the bulk of the theoretical literature to date, we make two
simplifications to make this general problem more approachable. First, we
consider only pairwise correlations; we do not take into account or discuss
the potential effects of higher-order correlations, which are more difficult
to estimate precisely from limited experimental data (see Refs.
\citep{cayco2015triplet,zylberberg2015input,montijn2016population} for examples
of recent studies of neural coding in the presence of higher-order
correlations). Second, we assume that the output of each individual neuron can
be represented by a scalar variable. This means, in particular, that we do not
consider temporal representations of information, such as those associated
with specific spike patterns. We think of the output of the neural population
as divided in successive time bins, and the activity of each neuron in each
time bin as defined by a single number (such as the spike count).
We examine the problem through the lens of
mathematical quantities that provide a characterization of the coding
performance independently of the choice of a putative decoder. Whenever
possible, we choose to explain theoretical results in terms of the mutual
(Shannon) information \citep{cover1999elements}. It quantifies 
information on a well-founded axiomatic basis, but has the disadvantage
that it is often difficult to calculate analytically. Besides its theoretical
foundation, our motivation in aligning various results in the framework of a
single mathematical `figure of merit' of the neural code is to provide as much
unity as possible to the discussion.

\section{Early views:\ homogeneous neural population with uniform noise
correlations}

Initial investigations suggested that noise correlation was detrimental to
neural coding. This conclusion was based on several (simplifying)
hypotheses:\ noise correlations were assumed to be positive, as suggested by
neural recordings, and uniform in a population of neurons with similar tuning
properties. Noise correlation was thus viewed as a `bug' in neural processing,
and possibly an unavoidable one due to the tight interconnections of neurons.

When we say \textquotedblleft noise correlation harms or benefits
coding,\textquotedblright\ we tacitly assume a comparison between a correlated
neural population and another neural population in all matters identical but
in which noise correlations have been removed. This independent population may
not be realizable in a real circuit due to interconnectedness of neurons, but
it provides a natural benchmark. Since we disregard higher-order correlations,
the comparison is between a correlated neural population and a neural
population in which neurons have identical mean responses and single-cell
variability around their mean responses, but in which pairwise correlations
are vanishing, i.e., a population of independent neurons with matched
single-cell response statistics. In practice, when analyzing data, there are
several ways to implement this comparison. A model-independent approach is to
create an artificial data set by shuffling recordings of individual neurons
among different experimental trials, in the population recording, so as to
retain single-cell statistics while eliminating the same-trial correlations.
If it is possible to fit a model to the population activity statistics, it is
also possible to compare this model to a parallel model in which the average
single-cell activity and noise variance are left unchanged while correlations 
of second and higher order are set to zero.

It is easy to see why positive noise correlation can be detrimental to coding
from the following simple model \citep{zohary_newsome_1994,bair_newsome_2001}.
Imagine that you want to discriminate two stimuli, A and B, from the output of
a population of $N$ neurons. For the sake of simplicity, we assume binary
neurons, i.e., the response of neuron $i$, $r_{i}$, can take the value $0$ or
$1$. If all the neurons in the population are identical in their response
properties, the state of the population is entirely characterized by the
number of active neurons,%
\begin{equation}
k=%
{\displaystyle\sum\limits_{i=1}^{N}}
r_{i}. \label{population-response}%
\end{equation}
On average over trials, $\left\langle k\right\rangle _{s}=Np\left(  s\right)
$, where the brackets, $\left\langle \cdot\right\rangle _{s}$, denote an
average over the distribution of population activity in the presence of
stimulus, $s$, and $p\left(  s\right)  $ is the probability that a neuron is
activated by the stimulus $s=$ A or B. From trial to trial, $k$ fluctuates
about this average quantity. The population output will discriminate the two
stimuli as long as the difference in the mean outputs, $N\left\vert p\left(
\text{A}\right)  -p\left(  \text{B}\right)  \right\vert $, is much larger than
the typical magnitude of these fluctuations,%
\begin{align}
\sqrt{\left\langle \left(  k-\left\langle k\right\rangle \right)
^{2}\right\rangle _{s}}  &  =\sqrt{%
{\displaystyle\sum\limits_{i,j=1}^{N}}
\left\langle \left[  r_{i}-p\left(  s\right)  \right]  \left[  r_{j}-p\left(
s\right)  \right]  \right\rangle _{s}}\nonumber\\
&  =\sqrt{N\left[  1+\left(  N-1\right)  c\left(  s\right)  \right]  p\left(
s\right)  \left[  1-p\left(  s\right)  \right]  },
\end{align}
where $c\left(  s\right)  $ is the pairwise correlation of two neurons in the
presence of stimulus $s$, defined as
\begin{equation}
c\left(  s\right)  =\frac{\left\langle \left[  r_{i}-p\left(  s\right)
\right]  \left[  r_{j}-p\left(  s\right)  \right]  \right\rangle _{s}}%
{\sqrt{\left\langle \left[  r_{i}-p\left(  s\right)  \right]  ^{2}%
\right\rangle _{s}\left\langle \left[  r_{j}-p\left(  s\right)  \right]
^{2}\right\rangle _{s}}}=\frac{\left\langle \left[  r_{i}-p\left(  s\right)
\right]  \left[  r_{j}-p\left(  s\right)  \right]  \right\rangle _{s}%
}{p\left(  s\right)  \left[  1-p\left(  s\right)  \right]  }.
\end{equation}
Assuming that the correlation does not depend much on the stimulus, $c\left(
\text{A}\right)  \approx c\left(  \text{B}\right)  \approx c$, we can define a
`signal-to-noise ratio' (SNR) that characterizes the faithfulness of the code
in discriminating the stimuli A and B, as%
\begin{equation}
\text{SNR}=\frac{N\left[  p\left(  \text{A}\right)  -p\left(  \text{B}\right)
\right]  ^{2}}{\left[  1+\left(  N-1\right)  c\right]  p\left(  1-p\right)  },
\label{SNR}%
\end{equation}
where $p$ lies somewhere between $p\left(  \text{A}\right)  $ and $p\left(
\text{B}\right)  $. This quantity is also the square of the `sensitivity
index' used in statistics and generally denoted by $d^{\prime}$.

The important point is that the SNR differs qualitatively for $c=0$ and
$c>0$ (Fig. 1A). In the case of \textit{independent} neurons ($c=0$), SNR
grows linearly and indefinitely with $N$. Each neuron added to the population
carries an incremental piece of information so that, roughly speaking, the
coding performance grows in proportion to the size of the population. This is
to be contrasted with the case of positively\textit{\ correlated} neurons
($c>0$): above a crossover size, $N^{\ast}\approx1/c$, positive correlation
limits the coding performance and the SNR saturates to a finite value at
larger population sizes (Fig. 1B). Each successive neuron added to a
growing population carries a decreasing amount of information, since its
variability is shared in part with that of all the other neurons in the
population. In large populations, the activity of an added neuron is
`dictated' by that of the other neurons and, hence, it does not provide any
incremental information.

\begin{figure}[tb]
\centering
\includegraphics[scale= 0.5]{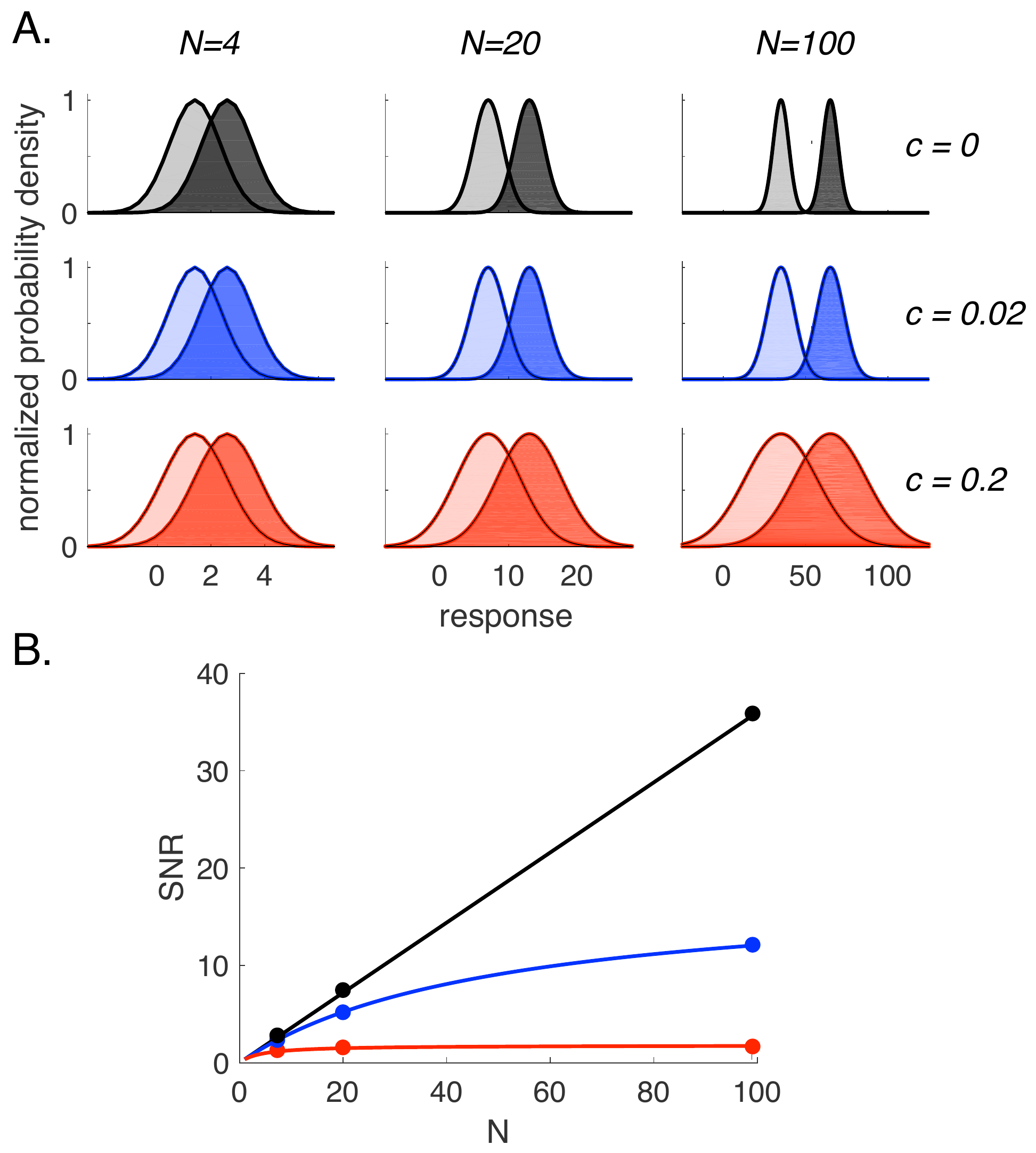}
\caption{{\bf Dependence of the signal-to-noise ratio on the number of
neurons ($N$) and the correlation strength ($c$).} \small{
{\bf A.} Probability densities of the
activity for several combinations of $N$ and $c$, in 
a homogenous population of neurons that
respond somewhat more strongly to stimulus $A$ (darker shaded
regions) than to stimulus $B$. For the purpose of illustration, neural responses are taken to be
Gaussian. The overlap between the two
distributions decreases steadily with $N$ in the case of independent neurons
($c=0$, black), 
but minimally in the case with $c=0.2$. \textbf{B.} Dependence of the
signal-to-noise ratio on $N$. Closed circles indicate parameter values as in
panel A.}}%
\end{figure}

Because of the form of the scaling with population size in Eq. (\ref{SNR}),
the effect of noise correlation can be strong even in relatively small
populations with modest values of correlation. For example, in the presence of
$10\%$ correlation ($c=0.1$, a typical value for cortical and retinal
neurons), noise correlation has an appreciable effect already in a population
of a few dozen neurons. For $N=100$, and assuming $p\left(  \text{A}\right)
-p\left(  \text{B}\right)  \approx p\approx0.5$, the signal-to-noise ratio
amounts to $9$, as opposed to $100$ for an independent population of neurons.
For $N=1000$, the signal-to-noise ratio grows to $10$, as opposed to $1000$
for an independent population of neurons. More generally, while the
signal-to-noise ratio increases by one unit for every independent neuron
added, in a correlated population it increases by an amount $\left(
1-c\right)  /\left(  1+Nc\right)  ^{2}$ when one neuron is added to a
population with $N$ neurons. With typical values of $c \approx 0.1$, this quantity drops rapidly to 
zero in populations with more than $100$ neurons.

There are at least two other ways of intuiting this result. The
signal-to-noise ratio acquires a factor of $N$ in its denominator because each
neuron shares a fraction of its variability with all the other neurons in the
population. As a result, any `error' committed by a neuron will be enhanced by
a factor of $N$, since neurons share their variability. Consequently, the
variability in the population response will be greatly enhanced. In other
words, positive correlation broadens the distribution of population responses.
Yet another way to think about this result is that positive correlation
induces neurons to respond similarly: it is as if positive correlation yields
a reduced `effective size' of the population, and, hence, suppresses the
coding capacity. In the extreme case of $100\%$ correlation ($c=1$), all
neurons in the population behave identically, and the population as a whole
cannot code for any more information than a single neuron does.

It is instructive to see how the conclusions obtained from the simple model
are reflected by a fundamental information theoretic quantity, the mutual
(Shannon) information. There are several equivalent ways to express the mutual
information; for our purposes we adopt the form%
\begin{equation}
I=\left\langle \sum_{r}P\left(  r\mid s\right)  \log\left(  \frac{P\left(
r\mid s\right)  }{\left\langle P\left(  r\mid s\right)  \right\rangle _{S}%
}\right)  \right\rangle _{S}, \label{mutual-information}%
\end{equation}
where $r\equiv\left(  r_{1},\ldots,r_{N}\right)  $ is the vector of population
response (or population activity), $s\in S$ denotes a stimulus (and $S$ is the
set of possible stimuli), and $\left\langle \cdot\right\rangle _{S}$ indicates
an average over all stimuli. In our simple model, there are two stimuli, $s=$
A or B, and $r$ labels the $2^{N}$ possible states of the population:%
\begin{equation}
r=\left(  n_{1},\ldots,n_{N}\right)  ,
\end{equation}
where $n_{i}=0$ if neuron $i$ is silent and $n_{i}=1$ if neuron $i$ is firing.
In this case, assuming that the two stimuli are
equiprobable,
\begin{equation}
\left\langle P\left(  r\mid s\right)  \right\rangle _{S}=\frac{1}{2}\left[
P\left(  r\mid\text{A}\right)  +P\left(  r\mid\text{B}\right)  \right]  ,
\end{equation}
we can rewrite Eq. (\ref{mutual-information}) as%
\begin{equation}
I=H-\frac{1}{2}\sum_{r}\left[  P\left(  r\mid\text{A}\right)  \log\left(
1+\frac{P\left(  r\mid\text{B}\right)  }{P\left(  r\mid\text{A}\right)
}\right)  +P\left(  r\mid\text{B}\right)  \log\left(  1+\frac{P\left(
r\mid\text{A}\right)  }{P\left(  r\mid\text{B}\right)  }\right)  \right]  ,
\label{mutual-information-AB}%
\end{equation}
where $H=\log\left(  2\right)  =1$ bit of information associated with the stimulus.

The second term on the right-hand-side of Eq. (\ref{mutual-information-AB}) is
referred to as the noise entropy and is a measure of the variability in the
neural response that is not due to the variability in the stimulus.  In other words, 
this term quantifies the amount of uninformative variability in the response. The noise
entropy is a sum of terms, each of which corresponds to a particular
realization of the population activity. From Eq. (\ref{mutual-information-AB}%
), it appears immediately that a given term vanishes if either of the
conditional response probabilities, $P\left(  r\mid\text{A}\right)  $ or
$P\left(  r\mid\text{B}\right)  $, vanishes; indeed, if a given stimulus
prevents a particular activity pattern, the latter is informative---it `codes'
for the other stimulus. Thus, the noise entropy grows as the overlap between
the two conditional response probabilities increases, and, correspondingly,
the mutual information is suppressed. If the overlap of the conditional
distributions does not decrease as $N$ increases, then the mutual information,
$I$, saturates and never reaches the stimulus entropy, $H$. In
this case, it is impossible to recover the full information about the stimulus
from the neural population response even in an infinitely large population, in
agreement with the picture from consideration of the signal-to-noise ratio.

\section{Broader views:\ coding in heterogeneous neural populations with
structured correlated noise}

Because neurons all come with identical properties in our simple model, there
is a single informative quantity: the total spike count, $k$ (Eq.
(\ref{population-response})). More generally, information is represented in a
higher-dimensional variable. Multiple studies
\citep{abbott_dayan_1999,sompolinsky_shamir_2001,wilke_eurich_2002,romo_salinas_2003,golledge_young_2003,averbeck_lee_2003,shamir_sompolinsky_2004,shamir_sompolinsky_2006,averbeck_lee_2006,averbeck_pouget_2006,josic_delarocha_2009,ecker2011effect,hu2013sign,silveira2013high,moreno2014information,franke2016structures,zylberberg2016direction}
have explored how noise correlation can affect the neural code in this case,
by exploiting higher-dimensional versions of the structure
illustrated in Fig. 1. The main---and important---departure
from our simple model was the generalization to heterogeneous neural
populations: the average single-neuron response to a given stimulus was
assumed to vary from neuron to neuron, and likewise pairwise correlations were
different from pair to pair. Most studies started with a set of `tuning
curves' (average response as a function of stimulus parameter) assigned to the
neurons in the population. Noise, including pairwise correlations, was either
estimated from neural recordings or posited on theoretical grounds. The
fidelity of the population code was then evaluated in terms of a chosen figure
of merit, such as the mutual information or a decoding error variance. The
results obtained thus depended on the specifics of the assumptions involved in
setting the forms of the tuning curves and of the noise model; to explore a
range of behaviors, the latter had to be varied. For example, many early
investigations used model neurons responding to a continuous stimulus with
broad tuning curves, and assumed noise models in which the pairwise
correlation depended only upon the tuning preferences of the two neurons in
the pair. Later studies included more sophisticated forms of heterogeneity and
dependences, such as the dependence of the pairwise correlation not only upon
the tuning preferences but also upon the stimulus itself.

To illustrate how coding depends on the manner in which neurons are
correlated, we take a more general but more qualitative approach. As stimulus
parameters are varied, the responses of the $N$ neurons in the population
trace out a hypersurface in the $N$-dimensional space of the population
responses (Fig. 2A). If the tuning curves are sufficiently smooth, this
hypersurface can be approximated locally by a hyperplane. (There exist
important examples in which this approximation is not valid
\citep{sreenivasan2011grid,blancomalerba}.) Single-trial population responses
depart from this hyperplane due to noise; the orientation of the hyperplane
and the geometry of the noise define $M$ `informative dimensions' or
`informative modes',%
\begin{equation}
m_{i}=%
{\displaystyle\sum\limits_{j=1}^{N}}
a_{ij}r_{j}, \label{informative-modes}%
\end{equation}
where $a_{ij}$ are numerical prefactors and $i=1,\ldots,M$. One can think of
these variables as chosen to maximize the mutual information with the stimulus
or to correspond to optimal decoding dimensions. If the noise is isotropic in
the $N$-dimensional space of population responses, then the informative
dimensions coincide with the hyperplane defined by the tuning curves; in
general, however, the `informative hyperplane' (defined by the coefficients
$a_{ij}$) and the `signal hyperplane' are distinct (Fig. 2C).

\begin{figure}[tbh]
\centering
\includegraphics[scale= 0.49]{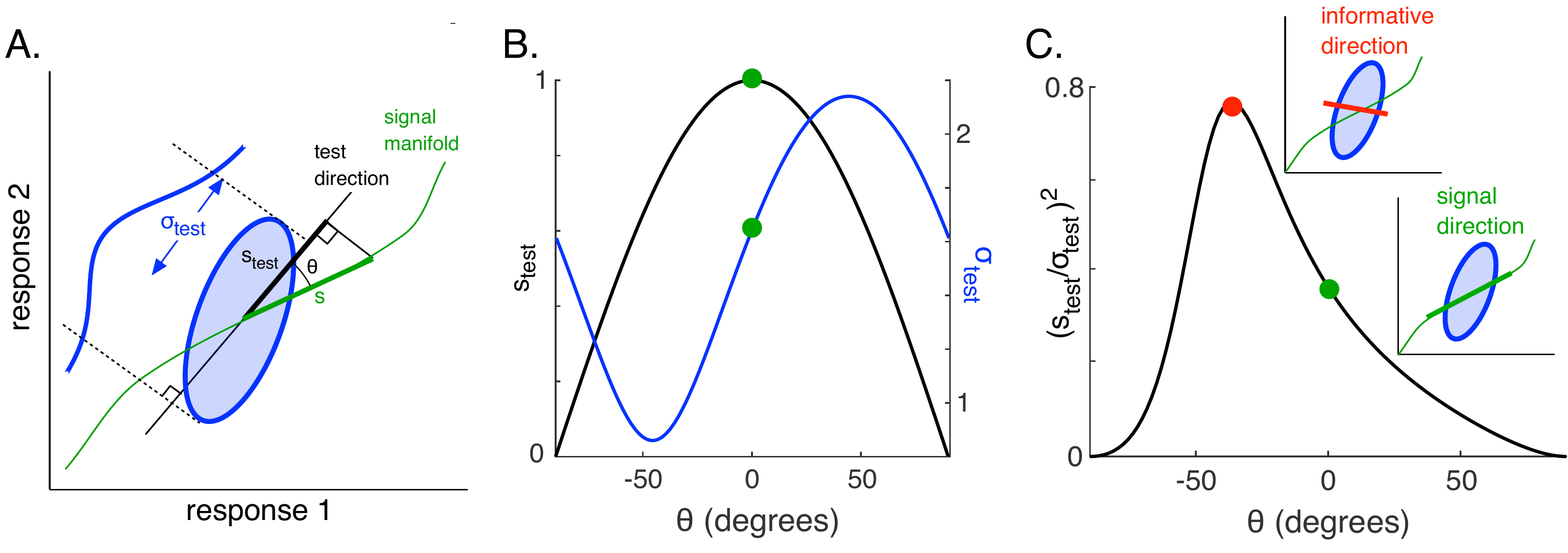}
\caption{{\bf Signal and noise together define the informative
direction for neural coding.} \small{{\bf A.} Two-neuron illustration of
quantities relevant for defining the signal-to-noise ratio in a specific direction.
The thin green line illustrates how the signal changes as the
stimulus is varied. The blue ellipse illustrates
the distribution of noisy responses corresponding to a mean
response located at one point along the green line. The signal-to-noise
ratio along a test direction (black) at an angle $\theta$ relative to the
signal direction can then be determined from the projection,
$s_{test}$, of the signal vector, $s$, and the
projection of the noise, $\sigma_{test}$ {\bf in the test direction}. \textbf{B.}
Signal and noise as a function of the angle between the test and signal
directions, for the situation depicted in
panel A. The green circles occur when the test and signal directions
are the same. \textbf{C.} The signal-to-noise ratio as a function of
$\theta$. Insets show the signal direction (green) and the informative
direction, the direction that maximizes the signal-to-noise ratio (red).}}%
\end{figure}

In the simplest and most commonly studied case of a one-dimensional stimulus,
the informative mode,%
\begin{equation}
m=%
{\displaystyle\sum\limits_{j=1}^{N}}
a_{j}r_{j}, \label{informative-mode}%
\end{equation}
lies along the vector with elements $a_{j}$. The quantity in Eq.
(\ref{informative-mode}) plays a role analogous to the spike count (Eq.
(\ref{population-response})) in our simple model. By analogy with the simple
model, the `strength of the signal' carried by the informative mode is
obtained by averaging over the noise, and grows linearly with the size of the
population: $\left\langle m\right\rangle \sim\mathcal{O}\left(  N\right) $.
%
How much information a mode represents depends also
upon the uncertainty of its value. Early studies considered cases in which
this uncertainty, as measured by its variance, grew either linearly or
quadratically with $N$
\citep{abbott_dayan_1999,sompolinsky_shamir_2001,wilke_eurich_2002,romo_salinas_2003}%
. If neurons are independent, the variance of the informative modes grows
linearly with the size of the population, so that each mode can represent
reliably up to about $\sqrt{N}$ different states of the stimuli. In this case,
the mutual information grows logarithmically in $N$. If, however, positive
correlation corrupts an informative mode, its typical amplitude grows
linearly with the size of the population and its variance grows quadratically;
in this case, the informative mode can represent reliably only $O\left(
1\right)  $ different states of the stimulus---i.e., the mutual
information saturates to a finite value smaller than the stimulus
entropy. More recent studies
\citep{ecker2011effect,hu2013sign,silveira2013high,franke2016structures,zylberberg2016direction}
(but see also Refs. \citep{abbott_dayan_1999,wilke_eurich_2002}) introduced
examples in which noise correlation may in fact suppress the variance of
informative modes relative to the independent case, thereby enhancing the
resolution of the code.

We can discuss these different cases by exploiting Eq. (\ref{informative-mode}%
), which we can rewrite as%
\begin{align}
m &  =%
{\displaystyle\sum\limits_{i=1}^{N}}
a_{i}\left\langle r_{i}\right\rangle +%
{\displaystyle\sum\limits_{i=1}^{N}}
a_{i}\eta_{i}\nonumber\\
&  \equiv\left\langle m\right\rangle +\mu,\label{informative-mode-expanded}%
\end{align}
where $\left\langle \cdot\right\rangle $ denotes an average over the noise and
the $\eta_{j}$s are $N$ correlated random variables with vanishing mean. The
second term, $\mu$, represents the uncertainty on the magnitude of the
informative mode and is the projection of the population noise along the
informative direction defined by the vector with elements $a_{j}$ (Fig. 2).
The informative mode can encode about as many different states of the stimulus
as the ratio between the first term, $\left\langle m\right\rangle $, and the
standard deviation of the second term, $\mu$, in Eq.
(\ref{informative-mode-expanded}). Its variance is calculated as%
\begin{align}
\left\langle \mu^{2}\right\rangle  &  =%
{\displaystyle\sum\limits_{i=1}^{N}}
a_{i}^{2}\left\langle \eta_{i}^{2}\right\rangle +%
{\displaystyle\sum\limits_{i=1}^{N}}
a_{i}%
{\displaystyle\sum\limits_{i\neq j}}
a_{j}\left\langle \eta_{i}\eta_{j}\right\rangle \nonumber\\
&  =%
{\displaystyle\sum\limits_{i=1}^{N}}
a_{i}\left(  a_{i}\left\langle \eta_{i}^{2}\right\rangle +Q_{i}\right)
,\label{informative-mode-var}%
\end{align}
where%
\begin{equation}
Q_{i}=%
{\displaystyle\sum\limits_{i\neq j}}
a_{j}\left\langle \eta_{i}\eta_{j}\right\rangle .
\end{equation}
The first term in Eq. (\ref{informative-mode-var}) represents the contribution
of independent neuron variance, and the second term represents the
contribution of correlated variability among neurons. Generically, $Q_{i}$ can
behave as a function of the population size in one of four ways, listed as
follows. 

(\textit{i}) $Q_{i}=0.$

(\textit{ii}) $Q_{i}\approx\pm a_{i}\left\langle \eta^{2}\right\rangle
\tilde{n}c.$

(\textit{iii}) $Q_{i}\approx a_{i}\left\langle \eta^{2}\right\rangle \tilde
{N}c.$

(\textit{iv}) $Q_{i}\approx-a_{i}\left\langle \eta^{2}\right\rangle \tilde
{N}c.$

Here, $\left\langle \eta^{2}\right\rangle $ corresponds to the typical scale of the
single-cell variance and $c\sim O\left(  1\right)  >0$ corresponds to
the typical scale of the (positive) pairwise noise correlation. The quantity
$Q_{i}$ also depends on an `effective population size' ($\tilde{n}$ or $\tilde{N}$) that
corresponds to the magnitude of the correlated noise mode relevant to coding.
Generically, $\tilde{n}\sim O\left(  1\right)  >0$ and $\tilde
{N}\sim O\left(  N\right)  >0$ are constants; in particular regimes,
$\tilde{N}$ can scale more weakly with $N$ (see below).
Without loss of generality, we exhibit a prefactor $a_{i}$ in these
expressions, for the sake of convenience given the form of Eq.
(\ref{informative-mode-var}). This form is natural, also, in the case of most
models considered in the literature, in which the total spike count in the
population is uninformative. For example, for neurons with broad tuning curves
that tile the stimulus space densely, so that the total spike count in the
population is roughly independent of the stimulus, the elements of the
informative vector sum to zero, i.e., $\sum_{i=1}^{N}a_{i}=0$. In a population
with uniform correlations \citep{abbott_dayan_1999,wilke_eurich_2002}, i.e.,
$\left\langle \eta_{i}\eta_{j}\right\rangle =\left\langle \eta^{2}%
\right\rangle c$\ for all $i,j$, the quantity $Q_{i}$ amounts to
$-a_{i}\left\langle \eta^{2}\right\rangle c$.

We can now organize the various results which appear in the literature among
these four categories:

\smallskip
\noindent\textit{1. Independence} (case \textit{i}). \noindent If neurons
are independent, $\left\langle \mu^{2}\right\rangle $ grows like $N$, so that
the informative mode can represent about $\left\langle m\right\rangle
/\sqrt{\left\langle \mu^{2}\right\rangle }\sim\sqrt{N}$ different states of
the stimulus.

\smallskip
\noindent\textit{2. Strongly detrimental noise correlation} (case
\textit{iii}). Early models
\citep{abbott_dayan_1999,sompolinsky_shamir_2001,wilke_eurich_2002}, assume
smooth, broad tuning curves, so that a given stimulus activates most neurons
in the population. As a result, the informative vector contains a macroscopic
fraction of non-vanishing elements, $a_{i}$, which vary slowly with $i$. If
the covariance of the noise, $\left\langle \eta_{i}\eta_{j}\right\rangle $,
also varies slowly with $j$ over the population, it can `interfere
constructively' with $a_{j}$, meaning that the noise is large in directions in
which the informative mode is also large. In this case, $Q_{i}$ grows like $N$
and $\left\langle \mu^{2}\right\rangle $ grows like $N^{2}$, and the
informative mode can represent only $\mathcal{O}\left(  1\right)  $ different
states of the stimulus. In other words, the performance of the code (as
measured, e.g., by the mutual information) saturates for large neural populations.

\smallskip
\noindent\textit{3. Weakly detrimental or weakly beneficial noise
correlation} (case \textit{ii}). Some early studies
\citep{abbott_dayan_1999,wilke_eurich_2002} noted that noise correlations that
are uniform over the population can lead to an improvement in the coding
performance. Indeed, if $\left\langle \eta_{i}\eta_{j}\right\rangle
=\left\langle \eta^{2}\right\rangle c$ is independent of $i$ and $j$, then
$Q_{i}=-a_{i}\left\langle \eta^{2}\right\rangle c$, and
\begin{equation}
\left\langle \mu^{2}\right\rangle =\sum_{i=1}^{N}a_{i}^{2}\left\langle
\eta^{2}\right\rangle \left(  1-c\right)  \sim\left(  1-c\right)
\mathcal{O}\left(  N\right)  .
\end{equation}
As a result, the informative mode can represent a number of different states
of the stimulus that depends upon the population size and the strength of the
noise correlation as $\sqrt{N}/\left(  1-c\right)  $. This moderate
improvement of the coding performance was obtained in studies that allowed
allowed for neuron-to-neuron variability in tuning curve properties
\citep{shamir_sompolinsky_2006,ecker2011effect}. This variability implied rapid
fluctuations of the elements of the informative vector, $a_{i}$, as a function
of the neuron index, $i$. If, by contrast, the noise covariance, $\left\langle
\eta_{i}\eta_{j}\right\rangle $, varies smoothly over the population, then the
`destructive interference' between these two terms yields, again, a linear
scaling of $Q_{i}$ with $N$, similar to the case of perfectly uniform
correlation. More generally, whether the noise correlation is strongly
detrimental or weakly detrimental/beneficial depends upon whether the
interference between informative vector and noise covariance is constructive
or destructive, respectively, over the population.

\smallskip
\noindent\textit{4. Strongly beneficial noise correlation} (case
\textit{iv}). Recent studies noted that the `destructive interference' between
the elements of the informative vector and the noise covariance can lead to an
appreciable suppression of the uncertainty
\citep{silveira2013high,franke2016structures,zylberberg2016direction}. This
occurs if $a_{j}$ and $\left\langle \eta_{i}\eta_{j}\right\rangle $ both vary
slowly as a function of $j$, over the population, but are, roughly speaking,
`out of phase': positive noise correlations are suppressed for neurons which
contribute to a greater degree to the `strength of the signal', and vice
versa. As a consequence, the quantity $Q_{i}$ becomes negative and scales with
$N$\thinspace\ and the variance of the noise is calculated as
\begin{equation}
\left\langle \mu^{2}\right\rangle \approx\alpha N\left\langle \eta
^{2}\right\rangle \left(  1-\tilde{N}c\right)  ,\label{strongly-beneficial}%
\end{equation}
where $\alpha$ is a positive number of $O\left(  1\right)
$ and $c$ is the typical scale of the (positive) pairwise
noise correlation as before. Generically, $\tilde{N}$ scales linearly
with $N$, so that uncertainty is strongly suppressed by noise
correlation, through the term $1-\tilde{N}c$.
The informative mode can then represent a number
of different states of the stimulus that depends upon the population size and
the strength of the noise correlation as $\sqrt{N}/\left(  1-\tilde{N}c\right)
$. The important point, here, is that the denominator is strongly
suppressed as a function of population size. This results, in particular, in
an appreciable enhancement of the coding performance when $\tilde{N}\sim O\left(
1/c\right)  $. 

The right-hand-side of Eq. (\ref{strongly-beneficial})
remains non-negative since the covariance of the noise is positive
semi-definite. In this formulation, we assume that the scale of the
correlation, characterized by $c$, is fixed; as $\tilde{N}$
increases, the covariance matrix becomes increasingly constrained by this
condition, and, depending on its structure, one or several small eigenvalues
may emerge. As these tend to zero, the informative vector rotates with respect
to the eigenvectors of the noise covariance matrix and, as a consequence, the
scaling of $\tilde{N}$ becomes weaker. This limiting regime in the
vicinity of a singular noise covariance matrix interpolates between the scalings in
cases (ii) and (iv), thereby allowing the term $1-\tilde{N}c$ in Eq.
(\ref{strongly-beneficial}) to remain non-negative. We return to the
discussion of the behavior of the informative direction as a function of the
structure of the noise in Sec. 6, where we provide further illustration in a
concrete model. 

The dependence of this boost upon the population size is a
signature of the collective effect at play here: in a correlated system, the
behavior of a neuron is affected by all \thinspace$N-1$ other neurons. For
$c\sim0.1$, as observed experimentally, the effect of correlation upon coding
is appreciable already in populations as small as tens or hundreds of neurons
\citep{silveira2013high}. A specific incarnation of this phenomenon occurs in a
model of broadly tuned neurons in which the dependence of the correlation
between a pair of neurons upon the difference in their tuning preference is
allowed to be non-monotonic
\citep{franke2016structures,zylberberg2016direction}.

The list just outlined catalogs the various ways in which noise correlation
can affect the coding of stimuli along an informative dimension by shaping the
variability in the population response. Our discussion of the `interference'
between elements of the informative vector and the noise covariances can be
seen as a generalization of the `sign rule' \citep{hu2013sign}, according to
which positive correlation is favorable in a pair of neurons with negative
signal correlation, and \textit{vice versa}.

There is one case, however, which was not covered: this is when there is no
informative dimension in the sense we discussed above. To be specific,
consider the case in which the average magnitude of the activity in the
`informative' dimension, $\left\langle m\right\rangle $, is independent of the
stimulus. Information about the stimulus can be encoded in the noise
itself: if correlation depends upon the stimulus, then different patterns of
population activity can discriminate stimuli. We return to this case in the
next section, in a more systematic treatment of the mutual information.

Finally, above we have considered only pairwise correlations. In the presence
of higher-order correlations, additional kinds of scalings occur. If real
neural systems are dominated by the strong co-activation of groups of neurons
corresponding to higher-order correlation, the analyses developed so far may
have a limited relevance to our understanding of population coding.

\section{A general approach to account for the impact of noise correlations on
mutual information}

Many of the studies of neural population coding to date have relied upon
specific models of neural populations, and have focused on one central
question: how does the coding performance scale with the number of neurons, in
particular in the limit of large populations? Moreover, most of these studies
quantified coding through the Fisher information. Other information-theoretic
quantities, such as the mutual (Shannon) information, are more fundamental
\citep{cover1999elements,brunel_nadal_1998,kang_sompolinsky_2001,wei2016mutual}
and avoid difficulties associated with the Fisher information
\citep{bethge2002optimal}. The Fisher information is local in stimulus space,
whereas the mutual information quantifies the accuracy of stimulus
representation over the entire stimulus space. The use of the Fisher
information also relies upon some restrictive assumptions, and yields only a
lower bound on the coding resolution, which may or may not be tight.

Coding in neural populations can be examined from a general perspective by
expressing the mutual information in a form that isolates the impact of
different types of correlation \citep{panzeri_rolls_1999,pola_panzeri_2003}. We
discuss the implications of this decomposition here; in App. B, we provide a
derivation of the decomposition, which follows and somewhat simplifies that in
Refs. \citep{panzeri_rolls_1999,pola_panzeri_2003}. The central result is a
reformulation of the mutual information as a sum of three terms:
\begin{equation}
I=I_{\text{independent}}+I_{\text{correlated}}^{\left(  1\right)
}+I_{\text{correlated}}^{\left(  2\right)  }.
\label{mutual-information-components}%
\end{equation}
Each of the terms can be expressed as a function of the joint probability,
$P\left(  r,s\right)  $, between stimulus, $s$, and population response, $r$,
and transformations of this joint probability, such as $P_{0}\left(  r\mid
s\right)  $, defined in Eq. (\ref{independent-conditional-probability}) which
denotes the conditional probability of the response in a population of
independent neurons with matched mean and variance. In App. B, we show that the three terms in
Eq. (\ref{mutual-information-components}) can be written as
\begin{equation}
I_{\text{independent}}\equiv\left\langle \sum_{r}P_{0}\left(  r\mid s\right)
\log\left(  \frac{P_{0}\left(  r\mid s\right)  }{\left\langle P_{0}\left(
r\mid s\right)  \right\rangle _{S}}\right)  \right\rangle _{S},
\end{equation}%
\begin{equation}
I_{\text{correlated}}^{\left(  1\right)  }=\left\langle \sum_{r}P\left(  r\mid
s\right)  \log\left(  \frac{P\left(  r\mid s\right)  /P_{0}\left(  r\mid
s\right)  }{\left\langle P\left(  r\mid s\right)  \right\rangle _{S}%
/\left\langle P_{0}\left(  r\mid s\right)  \right\rangle _{S}}\right)
\right\rangle _{S} \label{I-correlated-1}%
\end{equation}
and%
\begin{equation}
I_{\text{correlated}}^{\left(  2\right)  }=\left\langle \sum_{r}\left[
P\left(  r\mid s\right)  -P_{0}\left(  r\mid s\right)  \right]  \log\left(
\frac{\prod_{i=1}^{N}P\left(  r_{i}\right)  }{\left\langle P_{0}\left(  r\mid
s\right)  \right\rangle _{S}}\right)  \right\rangle _{S}.
\label{I-correlated-2}%
\end{equation}

The benefit of this reformulation of the mutual information is that each of
these three terms come with a transparent interpretation. The term
$I_{\text{independent}}$ represents the information carried by conditionally
independent neurons; indeed, if there is no noise correlation, $P\left(  r\mid
s\right)  =P_{0}\left(  r\mid s\right)  $, and both $I_{\text{correlated}%
}^{\left(  1\right)  }$ and $I_{\text{correlated}}^{\left(  2\right)  }$
vanish. Thus, $I_{\text{independent}}$ accounts for the amount of information
carried by the population which is not affected by noise correlations. In App.
B, we show that $\ I_{\text{independent}}$ can be broken down further to
isolate the impact of signal correlations.

The term $I_{\text{correlated}}^{\left(  1\right)  }$ is the formal analog to
the term $I_{\text{independent}}$, but where the information is carried by
noise correlations rather than by the structure of mean response. Thus, it
accounts for stimulus coding by the noise correlations themselves: the same
way differential firing rates characterize different stimuli, non-uniform
noise correlation can also specify the stimulus. Figure 3B illustrates an
example of the effects captured by this term in the simple case of a
two-neuron population that encodes a binary stimulus, $s=$ A or B. Here, the
mean responses to stimuli A and B are identical, yet the identity of the
stimulus can be inferred from the two-neuron response due to the differences
in noise correlation. Generalizations of this mechanism have been studied in various models of neural population coding \citep{shamir_sompolinsky_2004,josic_delarocha_2009,zylberberg2018role}, and stimulus-dependence of correlations has been proposed as supporting visual coding in direction-selective middle-temporal neurons in monkeys \citep{ponce2013stimulus}. 
As we show in App. B, both terms $I_{\text{independent}%
}$ and $I_{\text{correlated}}^{\left(  1\right)  }$ are non-negative; they
capture occurrences in which variations of mean response or noise correlations
as a function of stimulus are informative.

\begin{figure}[tbh]
\centering
\includegraphics[scale= 0.5]{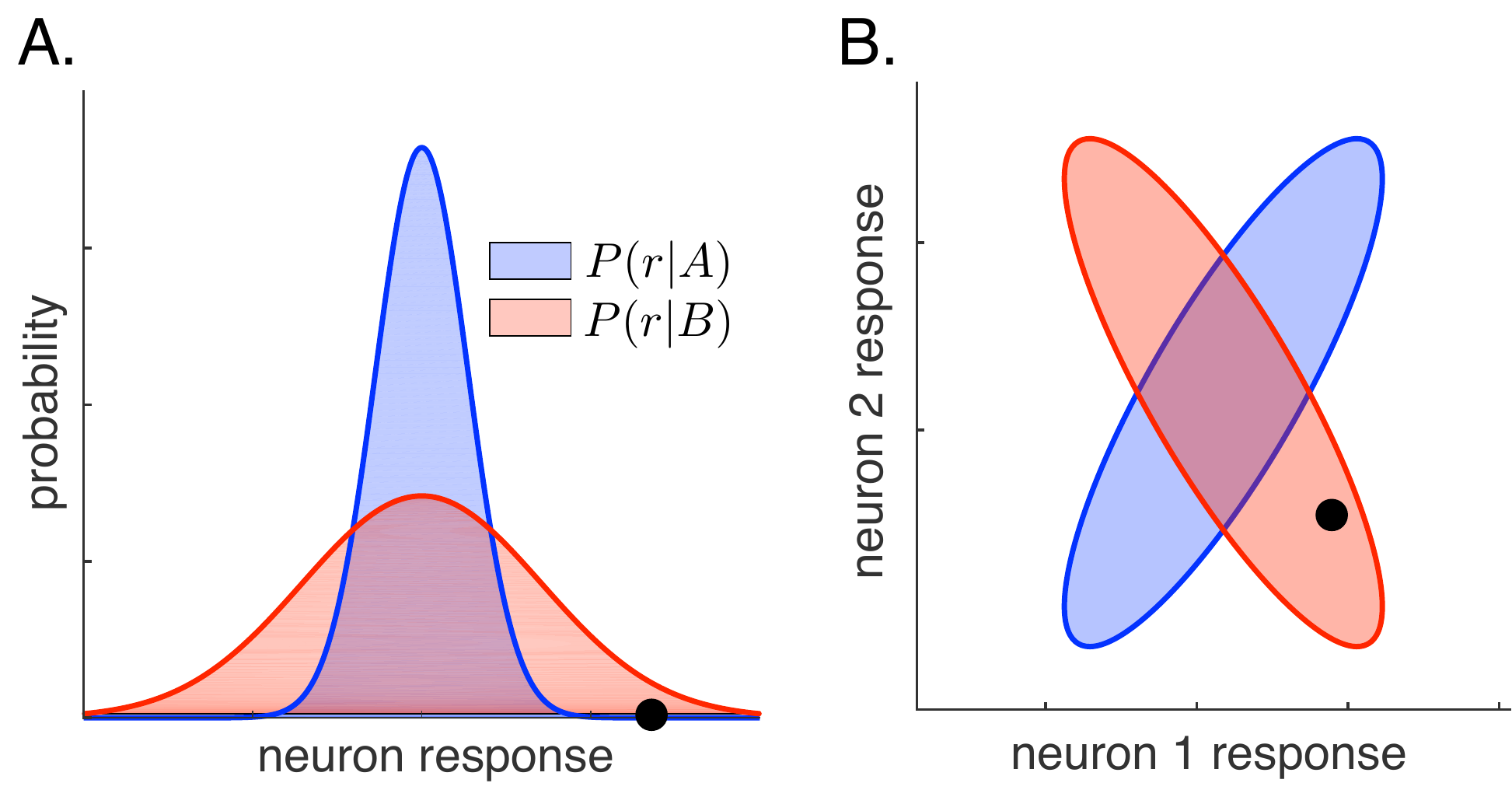}
\caption{{\bf Representation of information by stimulus-dependent noise.} \small{
{\bf A.} Illustration of a situation in which the variance of the response of
a single neuron encodes information about the stimulus identity. For example,
the noisy response denoted by the black circle is more likely to
be induced by stimulus $B$ than by stimulus $A$. \textbf{B.}
Illustration of a similar situation, in which the information is
encoded in correlations of the noise in the joint response of two neurons. 
In both panels A and B, the mean response is the same for stimulus A
and stimulus B, and, hence, uninformative about their identity.}}%
\end{figure}

Finally, $I_{\text{correlated}}^{\left(  2\right)  }$ represents the increment
or decrement of information due to the interplay between signal correlation
and noise correlation. From Eq. (\ref{I-correlated-2}), it is apparent that
$I_{\text{correlated}}^{\left(  2\right)  }$ vanishes if either noise
correlation is absent ($P\left(  r\mid s\right)  =P_{0}\left(  r\mid s\right)
$) or signal correlation is absent ($\left\langle P_{0}\left(  r\mid s\right)
\right\rangle _{S}=\prod_{i=1}^{N}P\left(  r_{i}\right)  $). Furthermore, and
unlike the other two components of the mutual information, this component can
be positive or negative. For a given stimulus, $s$, each population response,
$r$, yields a positive contribution if $\left[  P\left(  r\mid s\right)
-P_{0}\left(  r\mid s\right)  \right]  \times\left[  \prod_{i=1}^{N}P\left(
r_{i}\right)  -\left\langle P_{0}\left(  r\mid s\right)  \right\rangle
_{S}\right]  $ is positive, and a negative contribution otherwise. In other
words, the contribution to the information of a population response is
positive if noise correlation favors this population response while signal
correlation disfavors it, and \textit{vice versa}, and it is negative if both
noise and signal correlations either favor or disfavor the population
response. (When we say `favor' and `disfavor', as usual we are comparing the
case of a population of correlated neurons to the case of a population of
independent neurons.) The form of $I_{\text{correlated}}^{\left(  2\right)  }%
$\ in Eq. (\ref{I-correlated-2}) captures another generalized formulation of
the sign rule we mentioned in the previous section \citep{hu2013sign}. It also
illustrates in the language of mutual information, and without reference to
any specific model, the kind of `constructive versus destructive interference'
discussed in the previous section.

The reader might wonder about the merits of what may seem like a mere
mathematical exercise. Many studies today start from data and end in data, and
seek to make sense of data rather than to propose a theoretical framework. In
this context, we find it refreshing to examine the question from a more
abstract theoretical point of view, not tied to a specific model. It adds to
our understanding to be able to consider a same neural mechanisms from
multiple points of view. Having said this, we emphasize that the framework
just outlined can indeed be put to the task of analyzing data. It was recently
used, for example, to uncover the organization of assemblies of neurons with
redundant and synergistic coding of visual information in monkey cortex
\citep{nigam2019synergistic}.

\section{Information coding and the geometry of noise correlation}

The breakdown of the mutual information discussed in the previous section
teases apart the various contributions from signal and noise correlations, but
it provides neither a quantitative nor a geometrical view of how the
structures of signal and noise together impact coding. This section develops
such a geometrical view by revisiting the task of discriminating two stimuli,
A and B, discussed in Sec. 3.

To streamline the mathematical treatment, we consider a limiting case in which
the mutual information takes a simple form, namely, the case in which stimulus
A is presented with probability $\phi\ll1$. In this limit, the mutual
information can be written as
\begin{equation}
I=\phi\sum_{r}P\left(  r\mid\text{A}\right)  \ln\left(  \frac{P\left(
r\mid\text{A}\right)  }{P\left(  r\mid\text{B}\right)  }\right)
+\mathcal{O}\left(  \phi^{2}\right)  ,
\end{equation}
where $r$ is the vector of neural responses, $r^{T}=\left(  r_{1},\ldots
,r_{N}\right)  $. This approximation is valid when $\phi P\left(
r\mid\text{A}\right)  /P\left(  r\mid\text{B}\right)  \ll1$\thinspace\ i.e.,
we exclude cases in which there exist responses overwhelmingly more likely to
be elicited by stimulus A than by stimulus B. This condition is satisfied, in
particular, in the case of a demanding discrimination task, since in this case
the distributions $P\left(  r\mid\text{A}\right)  $ and $P\left(
r\mid\text{B}\right)  $ overlap considerably. If the noise is Gaussian, the
mutual information can be expressed in terms of the response mean and
covariance, as%
\begin{equation}
I=\phi\ln\left(  \frac{\operatorname*{Det}\left(  C_{\text{A}}\right)
}{\operatorname*{Det}\left(  C_{\text{B}}\right)  }\right)  +\frac{1}{2}%
\phi\operatorname*{Tr}\left(  C_{\text{A}}C_{\text{B}}^{-1}-\mathcal{I}%
\right)  +\frac{1}{2}\phi m^{T}C_{\text{B}}^{-1}m,
\label{mutual-information-Gaussian}%
\end{equation}
where $C_{s}$ ($s=\text{A,B}$) is the covariance of the noise in response to
stimulus $s$, $\mathcal{I}$ is the identity matrix, and $m$ is the `signal
vector' which, in this case, is simply the difference between the mean
responses to the two stimuli, i.e.,%
\begin{equation}
m_{i}\equiv\left\langle r_{i}\right\rangle _{\text{A}}-\left\langle
r_{i}\right\rangle _{\text{B}}.
\end{equation}
Here $i$ indicates the $i$th neuron, and $\left\langle \cdot\right\rangle
_{s}$ denotes the average over the conditional probability, $P\left(  r\mid
s\right)  $. To obtain Eq. (\ref{mutual-information-Gaussian}), we have further assumed that the spike
counts take large values, so that their discrete nature becomes unimportant
and the sum over population patterns can be replaced by an integral.

Equation (\ref{mutual-information-Gaussian}) can be interpreted particularly
transparently in the case in which the variances of the single-neuron
responses do not depend on the stimulus. In this case, the first two terms 
depend only upon the noise
correlation; they correspond to the term $I_{\text{correlated}}^{\left(
1\right)  }$ in Eq. (\ref{mutual-information-components}). The third term in
Eq. (\ref{mutual-information-Gaussian}) describes the interplay between signal
and noise correlation, and corresponds to the term $I_{\text{correlated}%
}^{\left(  2\right)  }$ in Eq. (\ref{mutual-information-Gaussian}). This term
is the formal equivalent to the so-called `linear Fisher information' used in
many earlier studies. It can also be viewed as the signal-to-noise ratio
discussed in Sec. 3, or the square of the sensitivity index, $d^{\prime}$.

We can examine the contribution of the interplay of signal and noise
correlation by comparing the third term in Eq.
(\ref{mutual-information-Gaussian}) for correlated and independent neurons,
again in the case in which the single-neuron variances do not depend on the
stimulus. For independent neurons, the mutual information reduces to
\begin{equation}
I_{0}\equiv\frac{1}{2}\phi m^{T}C_{0}^{-1}m,
\label{mutual-information-Gaussian-independent}%
\end{equation}
where $C_{0}$ is the diagonal matrix obtained from $C_{\text{B}}$ by setting
all off-diagonal elements to zero. If neurons are recorded from individually,
one has access only to the moments $m$ and $C_{0}$, and the mutual information
is estimated according to Eq. (\ref{mutual-information-Gaussian-independent}).
By contrast, when neurons in a population are recorded from simultaneously,
and the statistics of the population responses are fitted to a multivariate
Gaussian, then the mutual information is given by the richer Eq.
(\ref{mutual-information-Gaussian}).

Since the covariance matrix in Eq.
(\ref{mutual-information-Gaussian-independent}) is diagonal, the mutual
information for independent neurons can be rewritten as%
\begin{equation}
I_{0}=\frac{1}{2}\phi\sum_{i=1}^{N}\frac{m_{i}^{2}}{\sigma_{i}^{2}},
\label{trace-independent}%
\end{equation}
where $\sigma_{i}^{2}$ is the variance of the activity of neuron $i$. $I_{0}$
grows linearly as $N$ increases---a property of the limit of a rare stimulus
considered here. (Beyond a crossover size, the first-order expansion in $\phi$
breaks down, and, asymptotically, the mutual information increases
logarithmically in $N$). By analogy with Eq. (\ref{trace-independent}), a
natural way to calculate $I_{\text{correlated}}^{\left(  2\right)  }$, indeed
an approach followed by much of the literature (starting with Refs.
\citep{abbott_dayan_1999,sompolinsky_shamir_2001,wilke_eurich_2002}), is to
diagonalize the covariance matrix, to obtain%
\begin{equation}
I_{\text{correlated}}^{\left(  2\right)  }=\frac{1}{2}\phi\sum_{i=1}^{N}%
\frac{\tilde{m}_{i}^{2}}{\lambda_{i}}, \label{trace-correlated}%
\end{equation}
where $\tilde{m}_{i}$ are the elements of the vector $m$ in the new basis in
which $C_{\text{B}}$ is diagonal, and $\lambda_{i}$ is the $i$th eigenvalue of
the covariance matrix $C_{\text{B}}$. The argument is then that, if the
structure of pairwise correlations is such that the eigenvectors of the
covariance matrix (the `correlated modes') are not sparse and involve
contributions from a sizable fraction of the neurons in the population, then
the eigenvalues will scale with the population size. In this case, the sum in
Eq. (\ref{trace-correlated}) will yield a weaker scaling with $N$ than the sum
in Eq. (\ref{trace-independent}). In particular, if a few eigenvalues remain
small as $N$ increases, then these eigenvalues dominate the sum in Eq.
(\ref{trace-correlated}) and the latter asymptotes to a constant. In other
words, $I_{\text{correlated}}^{\left(  2\right)  }$ saturates to a finite
value in arbitrary large populations of neurons.

This approach is not entirely satisfactory because it is difficult to compare
Eq. (\ref{trace-independent}) and Eq. (\ref{trace-correlated}) since both the
numerator and the denominator differ. Indeed, the numerator in Eq.
(\ref{trace-correlated}) depends upon the signal vector as well as the
structure of the noise covariance. Furthermore, some of the eigenvalues in Eq.
(\ref{trace-correlated}) may take small values, and one may wonder what
dominates the sum: the larger terms associated with small eigenvalues or the
more numerous, smaller terms associated with larger eigenvalues. To resolve
these ambiguities, it is possible instead calculate an `information ratio'
that quantifies by how much noise correlations suppress or enhance coding as
compared to the case of an independent population of neurons
\citep{preprint2020}. This ratio can be
expressed in a compact form, as%
\begin{equation}
R_{I}\equiv\frac{I_{\text{correlated}}^{\left(  2\right)  }}{I_{0}}%
=\frac{\operatorname*{Det}\left(  \tilde{\chi}\right)  }{\operatorname*{Det}%
\left(  \chi\right)  }, \label{determinants}%
\end{equation}
where $\chi$ is the correlation matrix corresponding to the covariance matrix
$C_{\text{B}}$, and $\tilde{\chi}$ is the projection of $\chi$ on the $\left(
N-1\right)  $-dimensional subspace orthogonal to the modified signal vector,
$v$, with elements $v_{i}\equiv m_{i}/\sigma_{i}$. The information ratio
depends upon the spectra of the two matrices, $\chi$ and $\tilde{\chi}$. While
$\chi$ depends only upon noise correlation, $\tilde{\chi}$ incorporates an
interaction between the noise correlation the modified signal vector.

To intuit the behavior of the information ratio defined in Eq.
(\ref{determinants}), it is instructive to examine information coding with two
correlated neurons. The covariance of the noise reads%
\begin{equation}
C_{\text{B}}=\left(
\begin{array}
[c]{cc}%
\sigma_{1}^{2} & \sigma_{1}\sigma_{2}c\\
\sigma_{1}\sigma_{2}c & \sigma_{2}^{2}%
\end{array}
\right)  ,
\end{equation}
where $\sigma_{1}$ and $\sigma_{2}$ are the standard deviations in the
activities of the two neurons, and $c$ is the correlation of the noise. In
this simple case, the information ratio takes the form%
\begin{equation}
R_{I}=\frac{1-2c/\left(  \zeta+\zeta^{-1}\right)  }{1-c^{2}},
\end{equation}
where%
\begin{equation}
\zeta\equiv\frac{m_{1}}{\sigma_{1}}\left(  \frac{m_{2}}{\sigma_{2}}\right)
^{-1}.
\end{equation}
Since the parameter $\zeta$ can take any real value, inspection of the form of
the information ratio reveals that large volumes in the space of model
parameters yield $R_{I}>1$ and, conversely, $R_{I}<1$. Specifically, the
information ratio is larger than unity, i.e., noise correlation is beneficial
to information coding, when $c>2/\left(  \zeta+\zeta^{-1}\right)  $. This
relation, again, can be viewed as a generalization of the `sign rule':\ it
dictates how strong correlation ought to be to benefit coding as a function of
the signal vector and the single-neuron variances.

This simple example also helps shed light on the discussion in Sec.
4. There, we invoked an `informative dimension'. Similarly, here, we can ask
whether there is an especially informative dimension in the two-dimensional
space of the two-neuron population activity: in which direction should the
unit vector, $e$\thinspace\ point in order to maximize the mutual information
$I\left(  s;x\right)  $, where $s=$ A or B and $x\equiv e^{T}m=e_{1}%
m_{1}+e_{2}m_{2}$? This problem is solved easily, and the unit vector that
maximizes $I\left(  S;x\right)  $, call it $e^{\ast}$, can be expressed in
terms of the signal vector as well as the variances and correlation of the
pair of neurons. What is more interesting, though, is that the mutual
information $I\left(  S;x=e^{\ast T}m\right)  $ matches $I_{\text{correlated}%
}^{\left(  2\right)  }$ exactly: that is, the one-dimensional variable $x$
recovers the entirety of the useful information contained in the
two-dimensional activity of the neuron pair. The dimension defined by the
vector $e^{\ast}$ in the space of population activity is thus an `informative
dimension' in the sense of Sec. 4.

In general, the informative dimension does not align with the signal vector.
For example, in the limit of weak correlation, $\left\vert c\right\vert \ll1$,
and comparable variances, $\left\vert \sigma_{2}/\sigma_{1}-1\right\vert \ll
1$, the informative dimension is obtained by rotating the signal vector by an
angle $c\left(  m_{2}^{2}-m_{1}^{2}\right)  /m^{2}-2\left(  \sigma_{2}%
/\sigma_{1}-1\right)  m_{1}m_{2}/m^{2}$. The fidelity of coding depends upon
the noise along this direction, i.e., the variance of the projection of the
noise along $e^{\ast}$. A contrasting picture has been discussed in the
literature in recent years: a number of authors have argued that the fidelity
of coding depends, rather, on the presence of what they call `differential
correlations' \citep{moreno2014information,kohn2016correlations}. These are
taken to be present if the covariance matrix of the noise in the population
activity contains a component proportional to $mm^{T}$---i.e., a component
along the signal direction (Fig. 2A). The central conclusion from this line of
research is that differential correlations limit coding performance in that
they cause the information represented in the neural population to saturate
asymptotically, as $N\rightarrow\infty$.

A picture that emerges from the argument summarized in this section and
illustrated in Fig. 2 is that, if the covariance matrix contains small
eigenvalues, then the information represented in the neural population
(equivalently, the signal-to-noise ratio) can be large. This holds
\textit{even} if there is appreciable noise along the signal vector, provided
that the eigenvectors corresponding to the low-noise directions are
not orthogonal to the signal vector. More generally, figures of merit of the
coding performance of a population of neurons, such as the mutual information
or the signal-to-noise ratio, depend upon the full structure of the noise
covariance in relation to the signal vector, and not exclusively upon the
projection of the noise along the dimension defined by the signal vector.
Rather, in scenarios such as the ones discussed above, what matters is the
projection of the noise along an informative dimension in general distinct
from that along the signal vector. This is true in the case of finite values
of $N$ and in the asymptotic limit with $N\rightarrow\infty$. While some of
the mathematical statements can simplify in the asymptotic limit, this limit may
be far from natural for neural systems; for example, individual neurons may receive input
from a modest number of presynaptic neurons, and correlations in this collection of 
presynaptic neurons will shape signaling in the postsynaptic neuron.

\section{Future directions}

The discussion above aims at unifying various results in the literature using a
common metric for neural coding---the mutual information. We build intuition
about the impact of noise correlations on coding
by developing a geometrical picture of the structure of signal and noise. Our
goal is to highlight situations in which noise correlations are beneficial,
detrimental, or inconsequential for the
fidelity of the neural population code rather than to consider
specific examples that fall into one category or another. Below,
we summarize the assumptions that form the basis of our
discussion and we touch upon some of the open questions that
it raises.

Obstacles to understanding  coding in populations of
neurons arise largely from the high dimensionality of the problem. Experiments
by necessity probe a small subregion of the space of interesting stimuli, and the 
presence of response nonlinearities (such as adaptation) imply 
that insights gleaned from such experiments often 
do not generalize to all stimuli. The space of neural responses is similarly high dimensional, and is
impossible to probe completely in the finite duration of an experiment. 
But even an incomplete understanding of the role of noise correlations
in neural responses is helpful to guide experimental design. For instance, understanding
the neural code requires experiments that not only measure noise
correlations but also take into account their relation to
the encoded signal. Response variability 
may lie in a direction in which it impacts the encoded signal
minimally.  As an example, consider
a population of orientation-tuned V1 neurons. A change in stimulus orientation will increase activity in some neurons and decrease activity in others. Noise that produces correlated fluctuations
in firing rate that are uniform across the population will interact minimally with the signal
created by changes in orientation. 

Our discussion centers around correlations between pairs of neurons,
and neglects higher-order correlations. This is a
matter of practicality: pairwise correlations have been measured extensively
while we know much less about the properties of
higher-order correlations in real neural circuits. This is changing with the
availability of experimental approaches that allow  
a large number of cells to be monitored simultaneously. Theoretical frameworks that account for
geometrical relations between signal and higher-order noise correlations are
likely to play an important 
role in the development of new experimental protocols and methods of
data analyses, just as has been the case hitherto for pairwise correlations.

We chose to focus this review on how structures of
signal and noise interact from a statistical point of view, rather
than examining the neural
mechanisms that produce such structures. Single-cell
properties and the connectivity of real circuits constrain the structure of both signal and noise,
and a finer understanding of the connection between biological constraints and
network properties, on the one hand, and the statistics of population
response, on the other, will help interpret empirical observations 
\citep{doiron2016mechanics,rosenbaum2017spatial,huang2019circuit,trousdale2012impact,ocker2017linking,ostojic2009connectivity,mastrogiuseppe2018linking,schuessler2020dynamics,tannenbaum2017theory,goris2014partitioning,lin2015nature,pernice2011structure,pernice2018interpretation,delaRocha2007correlation,vidne2012commonnoise}.
A ubiquitous example is the divergence
of a common input into parallel
circuits, which can create both signal and noise correlations in those
circuits. This is just one of the many ways in which real circuit mechanisms shape signal and noise. Returning
to the picture of collective variables from statistical physics,
we can hope in the future to understand the relation between the
mechanisms that shape interactions between neurons, 
the impact of those interactions on the structure of signal and
noise, and how these combine to yield a representation of information in
neural populations.

Much of the literature in computational neuroscience focuses upon
population coding in the `thermodynamic limit' in which $N\longrightarrow
\infty$, that is, the limit of large populations of neurons. This is not, however, the 
only relevant limit. Individual neurons receive input from a finite set of other neurons; what matters for the
postsynaptic neuron is the representation of information in its finite
presynaptic pool. From a mathematical point of view, also, populations of
moderate sizes may be the relevant ones: for particular structures of pairwise
noise correlation, the condition $Nc\approx1$ defines a `strongly
correlated regime' in which stimuli can be encoded faithfully in the population
activity of tens or hundreds of neurons. And higher-order correlation can further enhance the coding
performance of `small' populations. Ultimately, we would like to relate the
coding accuracy in given populations of neurons to recorded perceptual acuity
and behavioral biases and variability. 

Historically, studies of noise correlation in neural activity were 
motivated precisely by questions of this type. A set of
early papers proposed that noise correlations relieved the necessity to
consider populations of more than a few hundred neurons, since the accuracy of
the encoded information saturated \citep{zohary_newsome_1994,bair_newsome_2001}%
. More recent studies have shown that noise correlations can be beneficial for
information coding 
\citep{averbeck_lee_2004,averbeck_lee_2006,montani2007role,graf2011decoding,lin2015nature,franke2016structures,zylberberg2016direction,montijn2016population}%
, and that when correlations are detrimental the encoded information saturates in much
larger populations of hundreds or thousands of neurons
\citep{bartolo2020information,rumyantsev2020fundamental}. One possible mechanism 
for this relative insensitivity of coding to noise is the alignment of
modes of strongly correlated noise away from the signal direction
\citep{bartolo2020information,rumyantsev2020fundamental}. At least one recent
(and unpublished) study \citep{stringer2019high}, however, suggests that mysteries still
lie ahead. The authors show that information encoded in mouse visual cortex does
not saturate in populations as large as 20000 neurons. Visual acuity inferred from activity
in these large populations outperforms mouse behavior by a factor of 10. This study
challenges our view of sensory coding. Mice may be able to
improve, through learning, their use of the encoded information; if
this is not the case, however, there may be fundamental reasons for which the brain
relinquishes the use of some of the information it has encoded. A satisfactory
understanding of information coding by neurons may only be possible through
the combined study of encoding and decoding in the brain, and behavior.

\pagebreak

\section*{Appendices}

\appendix

\section{Noise correlation \textit{versus} signal correlation}

When we consider the noisy response of a neural population to an ensemble of
stimuli, there are two possible averaging procedures: we can calculate
averages (moments) over the noise or over the ensemble of stimuli. Loosely
speaking, the former yields noise correlation while the latter yields signal
correlation. To be more specific, we consider a population of $N$ neurons and
we denote their outputs by $r_{1},\ldots,r_{N}$. The statistics of population
response is given by the conditional probability%
\begin{equation}
P\left(  r_{1},\ldots,r_{N}\mid s\right)  , \label{conditional-probability}%
\end{equation}
where $s$ refers to a stimulus chosen from a set of discrete stimuli or drawn
from a density over continuous stimuli. Noise correlations are non-vanishing
if%
\begin{equation}
P\left(  r_{1},\ldots,r_{N}\mid s\right)  \neq%
{\displaystyle\prod\limits_{i=1}^{N}}
P\left(  r_{i}\mid s\right)  \equiv P_{0}\left(  r_{1},\ldots,r_{N}\mid
s\right)  , \label{independent-conditional-probability}%
\end{equation}
where $P\left(  r_{i}\mid s\right)  $ is obtained from $P\left(  r_{1}%
,\ldots,r_{N}\mid s\right)  $ by averaging out all $r_{j}$, with $j\neq i$.
The probability function in Eq. (\ref{conditional-probability}) specifies the
\textit{noise correlations}; these characterize the population variability in
response to a given stimulus, and, hence, are themselves functions of the
stimulus. By contrast, signal correlation is a property if the statistics of
the population response over the ensemble or density of stimuli. Signal
correlations are obtained from the probability function%
\begin{equation}
\left\langle
{\displaystyle\prod\limits_{i=1}^{N}}
P\left(  r_{i}\mid s\right)  \right\rangle _{s}\equiv P_{0}\left(
r_{1},\ldots,r_{N}\right)  , \label{independent-probability}%
\end{equation}
where $\left\langle \cdot\right\rangle _{s}$ denotes an average over the
ensemble or density of stimuli. Signal correlations are non-vanishing if%
\begin{equation}
\left\langle
{\displaystyle\prod\limits_{i=1}^{N}}
P\left(  r_{i}\mid s\right)  \right\rangle _{s}\neq%
{\displaystyle\prod\limits_{i=1}^{N}}
\left\langle P\left(  r_{i}\mid s\right)  \right\rangle _{s}.
\end{equation}

Here, our focus will be on noise correlation and its effect upon sensory
coding. Whenever we mention `pairwise correlation' between two neurons labeled
by $i$ and $j$, we refer to the quantity calculated as%
\begin{equation}
c_{ij}=\frac{\left\langle \left(  r_{i}-\left\langle r_{i}\right\rangle
\right)  \left(  r_{j}-\left\langle r_{j}\right\rangle \right)  \right\rangle
}{\sqrt{\left\langle \left(  r_{i}-\left\langle r_{i}\right\rangle \right)
^{2}\right\rangle \left\langle \left(  r_{j}-\left\langle r_{j}\right\rangle
\right)  ^{2}\right\rangle }},
\end{equation}
where the average denoted by $\left\langle \cdot\right\rangle $ is weighed by
the conditional probability in Eq. (\ref{conditional-probability}). This
correlation coefficient follows the usual definition of a covariance
normalized by the corresponding standard deviations. By analogy, we can define
pairwise signal correlation as%
\begin{equation}
c_{ij}^{\text{signal}}=\frac{\left\langle \left(  \left\langle r_{i}%
\right\rangle -\left\langle \left\langle r_{i}\right\rangle \right\rangle
_{s}\right)  \left(  \left\langle r_{j}\right\rangle -\left\langle
\left\langle r_{j}\right\rangle \right\rangle _{s}\right)  \right\rangle _{s}%
}{\sqrt{\left\langle \left(  \left\langle r_{i}\right\rangle -\left\langle
\left\langle r_{i}\right\rangle \right\rangle _{s}\right)  ^{2}\right\rangle
_{s}\left\langle \left(  \left\langle r_{j}\right\rangle -\left\langle
\left\langle r_{j}\right\rangle \right\rangle _{s}\right)  ^{2}\right\rangle
_{s}}},
\end{equation}
where the average denoted by $\left\langle \cdot\right\rangle _{s}$ is weighed
by the `prior probability' over stimuli, $P\left(  s\right)  $.

\section{Breakdown of the mutual information in terms of signal and noise}

We follow Refs. \citep{panzeri_rolls_1999,pola_panzeri_2003} which introduce a
breakdown of the mutual information in several terms that exhibit the various
ways in which noise correlations may influence the coding performance. We
spell out a somewhat modified derivation, here, as the latter is possibly a
more direct one. We reformulate the mutual information (Eq.
(\ref{mutual-information})) in a form that emphasizes the contributions of
signal and noise correlations.

To achieve this, we invoke the independent (marginalized) probabilities
defined in Eqs. (\ref{independent-conditional-probability}) and
(\ref{independent-probability}), and rewrite the mutual information in terms
of the independent conditional probability, $P_{0}\left(  r\mid s\right)  $,
and ratios that carry the contribution of correlations, $P\left(  r\mid
s\right)  /P_{0}\left(  r\mid s\right)  $ and $\left\langle P\left(  r\mid
s\right)  \right\rangle _{S}/\left\langle P_{0}\left(  r\mid s\right)
\right\rangle _{S}$, as%

\begin{equation}
I=\left\langle \sum_{r}P\left(  r\mid s\right)  \left[  \log\left(
\frac{P\left(  r\mid s\right)  /P_{0}\left(  r\mid s\right)  }{\left\langle
P\left(  r\mid s\right)  \right\rangle _{S}/\left\langle P_{0}\left(  r\mid
s\right)  \right\rangle _{S}}\right)  +\log\left(  \frac{P_{0}\left(  r\mid
s\right)  }{\left\langle P_{0}\left(  r\mid s\right)  \right\rangle _{S}%
}\right)  \right]  \right\rangle _{S}.
\end{equation}
We further separate independent probabilities from correlated ones by adding a
subtracting to this quantity the mutual information corresponding to an
independent population of neurons, i.e., the term%
\begin{equation}
I_{\text{independent}}\equiv\left\langle \sum_{r}P_{0}\left(  r\mid s\right)
\log\left(  \frac{P_{0}\left(  r\mid s\right)  }{\left\langle P_{0}\left(
r\mid s\right)  \right\rangle _{S}}\right)  \right\rangle _{S}.
\end{equation}
This manipulation allows us to to obtain the form in Eq.
(\ref{mutual-information-components}), i.e.,
\begin{equation}
I=I_{\text{independent}}+I_{\text{correlated}}^{\left(  1\right)
}+I_{\text{correlated}}^{\left(  2\right)  },
\end{equation}
where $I_{\text{correlated}}^{\left(  1\right)  }$ is defined in Eq.
(\ref{I-correlated-1}) and%
\begin{equation}
I_{\text{correlated}}^{\left(  2\right)  }=\left\langle \sum_{r}\left[
P\left(  r\mid s\right)  -P_{0}\left(  r\mid s\right)  \right]  \log\left(
\frac{P_{0}\left(  r\mid s\right)  }{\left\langle P_{0}\left(  r\mid s\right)
\right\rangle _{S}}\right)  \right\rangle _{S}.
\label{I-correlated-2-intermediate}%
\end{equation}

The quantity $I_{\text{independent}}$ represents the information carried by
conditionally independent neurons; indeed, if there is no noise correlation,
$P\left(  r\mid s\right)  =P_{0}\left(  r\mid s\right)  $, and both
$I_{\text{correlated}}^{\left(  1\right)  }$ and $I_{\text{correlated}%
}^{\left(  2\right)  }$ vanish. It can be broken down further, to extract the
contribution from signal correlation, by bringing in the single-cell
marginalized probabilities,
\begin{equation}
P\left(  r_{i}\right)  \equiv\left\langle P_{0}\left(  r_{i}\mid s\right)
\right\rangle _{s}.
\end{equation}
With these, we can rewrite $I_{\text{independent}}$ as%
\begin{align}
I_{\text{independent}}  &  =\left\langle \sum_{r}P_{0}\left(  r\mid s\right)
\log\left(  \frac{P_{0}\left(  r\mid s\right)  }{\prod\nolimits_{i=1}%
^{N}P\left(  r_{i}\right)  }\right)  \right\rangle _{S}-\left\langle \sum
_{r}P_{0}\left(  r\mid s\right)  \log\left(  \frac{\left\langle P_{0}\left(
r\mid s\right)  \right\rangle _{S}}{\prod\nolimits_{i=1}^{N}P\left(
r_{i}\right)  }\right)  \right\rangle _{S}\nonumber\\
&  =I_{\text{independent}}^{\left(  1\right)  }-I_{\text{independent}%
}^{\left(  2\right)  },
\end{align}
with%
\begin{equation}
I_{\text{independent}}^{\left(  1\right)  }=\sum_{i=1}^{N}\left\langle
\sum_{r}P_{0}\left(  r_{i}\mid s\right)  \log\left(  \frac{P_{0}\left(
r_{i}\mid s\right)  }{\left\langle P_{0}\left(  r_{i}\mid s\right)
\right\rangle _{s}}\right)  \right\rangle _{s}
\label{mutual-information-independent}%
\end{equation}
and%
\begin{equation}
I_{\text{independent}}^{\left(  2\right)  }=\sum_{r}\left\langle P_{0}\left(
r\mid s\right)  \right\rangle _{S}\log\left(  \frac{\left\langle P_{0}\left(
r\mid s\right)  \right\rangle _{S}}{\prod_{i=1}^{N}\left\langle P_{0}\left(
r_{i}\mid s\right)  \right\rangle _{S}}\right)  .
\end{equation}
By comparing the form of Eq. (\ref{mutual-information-independent}) with that
of Eq. (\ref{mutual-information}), we see that it expresses the sum over the
information carried by $N$ independent neurons. Since $I_{\text{independent}%
}^{\left(  2\right)  }$ vanishes in the absence of signal correlation,
$I_{\text{independent}}^{\left(  1\right)  }$ amounts to the total mutual
information if both signal and noise correlations vanish. The quantity
$I_{\text{independent}}^{\left(  2\right)  }$ thus represents the loss of
information due signal correlation; indeed, $I_{\text{independent}}^{\left(
2\right)  }$ is nothing but the difference between the entropy of the
marginalized \textit{independent} distribution, $\prod_{i=1}^{N}\left\langle
P_{0}\left(  r_{i}\mid s\right)  \right\rangle _{S}$, and the entropy of the
marginalized \textit{correlated} distribution, $\left\langle P_{0}\left(
r\mid s\right)  \right\rangle _{S}$,%
\begin{equation}
I_{\text{independent}}^{\left(  2\right)  }=-\sum_{r}\prod_{i=1}%
^{N}\left\langle P_{0}\left(  r_{i}\mid s\right)  \right\rangle _{S}%
\log\left(
{\displaystyle\prod\limits_{i=1}^{N}}
\left\langle P_{0}\left(  r_{i}\mid s\right)  \right\rangle _{S}\right)
+\sum_{r}\left\langle P_{0}\left(  r\mid s\right)  \right\rangle _{S}%
\log\left(  \left\langle P_{0}\left(  r\mid s\right)  \right\rangle
_{S}\right)  ,
\end{equation}
and, as such, is non-negative.

The quantity $I_{\text{correlated}}^{\left(  1\right)  }$ represents the
information carried by noise correlation. By rewriting the logarithm in Eq.
(\ref{I-correlated-1}) as%
\begin{equation}
\log\left(  \frac{P\left(  r\mid s\right)  /P_{0}\left(  r\mid s\right)
}{\left\langle P\left(  r\mid s\right)  /P_{0}\left(  r\mid s\right)
\right\rangle _{S}}\right)  +\log\left(  \frac{\left\langle P\left(  r\mid
s\right)  /P_{0}\left(  r\mid s\right)  \right\rangle _{S}}{\left\langle
P\left(  r\mid s\right)  \right\rangle _{S}/\left\langle P_{0}\left(  r\mid
s\right)  \right\rangle _{S}}\right)  ,
\end{equation}
then using convexity and Cauchy-Schwarz inequalities, one can show that
$I_{\text{correlated}}^{\left(  1\right)  }$ is non-negative. It vanishes if
the ratio $P\left(  r\mid s\right)  /P_{0}\left(  r\mid s\right)  $ is
independent of the stimulus. Thus, $I_{\text{correlated}}^{\left(  1\right)
}$ accounts for stimulus coding by the values of the noise correlations
themselves (Fig. 3): the same way differential firing rates characterize
different stimuli, non-uniform noise correlation can also specify the stimulus.

Finally, the quantity $I_{\text{correlated}}^{\left(  2\right)  }$ represents
the increment or decrement of information due to the interplay between signal
correlation and noise correlation. This appears if we rewrite its expression
to emphasize the contribution of signal correlation, by replacing the ratio%
\begin{equation}
\frac{P_{0}\left(  r\mid s\right)  }{\left\langle P_{0}\left(  r\mid s\right)
\right\rangle _{S}}\text{ \ by the product \ }\frac{\prod_{i=1}^{N}P\left(
r_{i}\right)  }{\left\langle P_{0}\left(  r\mid s\right)  \right\rangle _{S}%
}\cdot\frac{P_{0}\left(  r\mid s\right)  }{\prod_{i=1}^{N}P\left(
r_{i}\right)  }.
\end{equation}
We then rewrite $I_{\text{correlated}}^{\left(  2\right)  }$ as%
\begin{align}
I_{\text{correlated}}^{\left(  2\right)  }  &  =\left\langle \sum_{r}\left[
P\left(  r\mid s\right)  -P_{0}\left(  r\mid s\right)  \right]  \log\left(
\frac{\prod_{i=1}^{N}P\left(  r_{i}\right)  }{\left\langle P_{0}\left(  r\mid
s\right)  \right\rangle _{S}}\right)  \right\rangle _{S}\nonumber\\
&  +\left\langle \sum_{r}\left[  P\left(  r\mid s\right)  -P_{0}\left(  r\mid
s\right)  \right]  \log\left(  \frac{P_{0}\left(  r\mid s\right)  }%
{\prod_{i=1}^{N}P\left(  r_{i}\right)  }\right)  \right\rangle _{S},
\end{align}
but the second term in fact yields a vanishing contribution:%
\begin{align}
&  \left\langle \sum_{r}\left[  P\left(  r\mid s\right)  -P_{0}\left(  r\mid
s\right)  \right]  \log\left(  \frac{P_{0}\left(  r\mid s\right)  }%
{\prod_{i=1}^{N}P\left(  r_{i}\right)  }\right)  \right\rangle _{S}\nonumber\\
&  =\sum_{i=1}^{N}\left\langle \sum_{r}\left[  P\left(  r\mid s\right)
-P_{0}\left(  r\mid s\right)  \right]  \log\left(  \frac{P_{0}\left(
r_{i}\mid s\right)  }{\prod_{i=1}^{N}P\left(  r_{i}\right)  }\right)
\right\rangle _{S}\nonumber\\
&  =\sum_{i=1}^{N}\left\langle \sum_{r_{i}}\left[  P\left(  r_{i}\mid
s\right)  -P_{0}\left(  r_{i}\mid s\right)  \right]  \log\left(  \frac
{P_{0}\left(  r_{i}\mid s\right)  }{P\left(  r_{i}\right)  }\right)
\right\rangle _{S}\nonumber\\
&  =0,
\end{align}
since $P\left(  r_{i}\mid s\right)  =P_{0}\left(  r_{i}\mid s\right)  $.
Hence, $I_{\text{correlated}}^{\left(  2\right)  }$ (Eq.
(\ref{I-correlated-2-intermediate})) can be written in the simpler form given
in Eq. (\ref{I-correlated-2}).

\section*{DISCLOSURE STATEMENT}
The authors are not aware of any affiliations, memberships, funding, or financial holdings that
might be perceived as affecting the objectivity of this review. 

\section*{ACKNOWLEDGMENTS}
Support was provided by the CNRS through UMR 8023, the SNSF Sinergia Project CRSII5\_173728, and the National Institute of Health (EY028111 and EY028542).

\vfill\eject

\bibliographystyle{ar-style1}
\setcitestyle{authoryear, open={(},close={)}}
\bibliography{References-CorrelationsAndCoding}

\end{document}